\documentclass[12pt]{iopart}
\usepackage{iopams}
\usepackage{amssymb}
\begin{document}
\title{Gazeau-Klauder type coherent states for hypergeometric type operators}
\author{Nicolae Cotfas}
\address{Faculty of Physics, University of Bucharest,
PO Box 76-54, Post Office 76, Bucharest, Romania,\\ 
E-mail address: ncotfas@yahoo.com } 
\begin{abstract}
The hypergeometric type operators are shape invariant, and a factorization into a product of
first order differential operators can be explicitely described in the general case. Some additional shape invariant
operators depending on several parameters are defined in a natural way by starting from this general 
factorization. The mathematical properties of the eigenfunctions and eigenvalues of the operators thus 
obtained depend on the values of the involved parameters. 
We study the parameter dependence of orthogonality, square integrability and of the monotony of eigenvalue sequence. The obtained results allow us
to define certain systems of Gazeau-Klauder coherent states and to describe some of their properties.
Our systematic study recovers a number of well-known results in a natural unified way and also leads to
new findings. 
\end{abstract}
\maketitle

\section{Introduction}
Many problems in quantum mechanics and
mathematical physics lead to equations of the type
\begin{equation}\label{hypeq}
\sigma (s)y''(s)+\tau (s)y'(s)+\lambda y(s)=0 
\end{equation}
where $\sigma (s)$ and $\tau (s)$ are polynomials of at most second
and first degree, respectively, and $\lambda $ is a constant. 
These equations are usually called {\em equations of hypergeometric
type} \cite{NSU}, and each of them can be reduced to the self-adjoint form 
\begin{equation}
[\sigma (s)\varrho (s)y'(s)]'+\lambda \varrho (s)y(s)=0 
\end{equation}
by choosing a function $\varrho $ such that 
$(\sigma \varrho )'=\tau \varrho $.
The equation (\ref{hypeq}) is usually considered on an interval $(a,b)$,
chosen such that 
\begin{equation}\label{cond}
\begin{array}{r}
\sigma (s)>0\qquad {\rm for\ all}\quad s\in (a,b)\\
\varrho (s)>0\qquad {\rm for\ all}\quad s\in (a,b)\\
\lim_{s\rightarrow a}\sigma (s)\varrho (s)
=\lim_{s\rightarrow b}\sigma (s)\varrho (s)=0.
\end{array}
\end{equation}
Since the form of the equation (\ref{hypeq}) is invariant under a 
change of variable $s\mapsto cs+d$, it is sufficient to analyze the cases
presented in table 1.
Some restrictions are imposed on $\alpha $ and $\beta $ in
order that the interval $(a,b)$ exists. One can remark that the second derivative 
$\sigma ''$ of $\sigma $ belongs to  $\{ 0,\, -2,\, 2\}$.

\begin{table}[htbp]
\caption{The main cases}
\begin{center}
\begin{tabular}{cclll}
\hline
$\sigma (s)$ & $\tau (s)$  & $\varrho (s)$  & $(a,b) $ & $\alpha ,\beta $ \\
\hline \hline
$1$ & $\alpha s+\beta $ & ${\rm e}^{\alpha s^2/2+\beta s}$ & $(-\infty ,\infty )$\quad \mbox{} & $\alpha <0$\\
$s$ & $\alpha s+\beta $ & $s^{\beta -1} {\rm e}^{\alpha s}$ & $(0,\infty )$ & $\alpha <0$, $\beta >0$ \\ 
$1-s^2$  & $\alpha s+\beta $ & $(1+s)^{-(\alpha -\beta )/2-1}
(1-s)^{-(\alpha +\beta )/2-1}$  & $(-1,1)$ & $\alpha <\beta <-\alpha $\\
$s^2-1$  & $\alpha s+\beta $ & $(s+1)^{(\alpha -\beta )/2-1}
(s-1)^{(\alpha +\beta )/2-1}$  & $(1,\infty )$ & $-\beta <\alpha <0$\\
$s^2$  & $\alpha s+\beta $ & $s^{\alpha -2}{\rm e}^{-\beta /s}$  &
$(0,\infty )$ & $\alpha <0$, $\beta >0$\\
$s^2+1$  & \mbox{}\quad $\alpha s+\beta $\quad \mbox{} & $(1+s^2)^{\alpha /2-1}{\rm e}^{\beta \arctan s}$ 
 & $(-\infty ,\infty )$ & $\alpha <0$ \\
\hline
\end{tabular}
\end{center}
\end{table}

In the paper \cite{C1} our aim was to analyze together in a unified formalism all the 
cases arising by imposing the conditions (\ref{cond}). The systems of polynomials $\Phi _l$
obtained in this way can be expressed in terms of the classical orthogonal polynomials, 
but in some cases we have to consider the classical polynomials for complex values of parameters 
or outside the interval where they are orthogonal.
The literature discussing special function theory and its application to mathematical
and theoretical physics is vast, and there is a multitude of different conventions
concerning the definition of functions. A unified approach 
is not possible without a unified definition. The associated special functions we define as
\begin{equation}
 \Phi _{l,m}(s)=\kappa ^m(s)\, \frac{{\rm d}^m}{{\rm d}s^m}\Phi _l(s)\qquad {\rm with}\qquad 
\kappa (s)\!=\!\sqrt{\sigma (s)}
\end{equation}
are eigenfunctions of the hypergeometric type operators 
\begin{equation} \fl
H_m =-\sigma (s) \frac{d^2}{ds^2}-\tau (s) \frac{d}{ds}
+\frac{m(m-2)}{4}\frac{(\sigma '(s))^2}{\sigma (s)}   
 + \frac{m\tau (s)}{2}\frac{\sigma '(s)}{\sigma (s)}
-\frac{1}{2}m(m-2)\sigma ''(s)-m\tau '(s)  
\end{equation}
depending on parameters $\alpha $, $\beta $ and $m$. The operators $H_m$ satisfy the relations
\begin{equation} \label{factoriz}
\begin{array}{ll}
H_m-\lambda _m=A_m^+A_m\qquad &  H_mA_m^+=A_m^+H_{m+1}\\
H_{m+1}\!-\!\lambda _m=A_mA_m^+  \qquad &  A_mH_m=H_{m+1}A_m
\end{array} 
\end{equation}
where $\lambda _m\!=\!-\frac{\sigma ''}{2}m(m\!-\!1)\!-\!\tau 'm$ and 
\begin{equation} 
A_m\!=\!\kappa (s)\frac{d}{ds}\!-\!m\kappa '(s)\qquad \qquad 
A_m^+\!=\!-\kappa (s)\frac{d}{ds}\!-\!\frac{\tau (s)}{\kappa (s)}\!-\!(m\!-\!1)\kappa '(s).
\end{equation}
In the particular case when $\alpha $ and $\beta $ are such that $\varrho $ is a power of $\sigma $
the factorization (\ref{factoriz}) allows one to prove that the operators
\begin{equation}
\tilde H_m=H_m-\delta \, \kappa '(s)
\end{equation}
depending on an additional parameter $\delta $ satisfy some relations similar to (\ref{factoriz}). 
Certain results concerning the operators $\tilde H_m$ can be found in \cite{C2,JF}. 
The mathematical properties of the operators $\tilde H_m$
depend on the involved parameters, and our main purpose is to investigate this dependence. Our approach is 
based on the similitude existing between $H_m$ and $\tilde H_m$. Therefore, we firstly review certain properties of
$H_m$ in a form suitable for our purpose. If the involved parameters satisfy certain restrictions,
some Gazeau-Klauder systems of coherent states can be associated in a natural way. We consider 
the notion of Gazeau-Klauder coherent states in a larger sense. The measure we use is not positive in all
the considered cases. In the resolution of identity, certain states bring a positive contribution while the others bring
a negative contribution.\\[3mm]
It is well-known that the hypergeometric type operators $H_m$ and $\tilde H_m$ are directly related to 
some exactly solvable Schrodinger type equations including a large number of potentials 
(shifted oscillator, three-dimensional oscillator, P\"{o}schl-Teller, generalized P\"{o}schl-Teller, 
Morse, Scarf hyperbolic, Coulomb, trigonometric Rosen-Morse, Eckart, hyperbolic Rosen-Morse) and 
their supersymmetric parteners. Most of the mathematical formulae occurring in the study of the exactly 
solvable quantum systems follow from a small number of formulae concerning the hypergeometric type operators. 
Particularly, almost any factorization used in quantum mechanics for a Schrodinger operator 
follows through a change of variable \cite{IH,CKS,C2} from the explicitely described factorization (\ref{factoriz}).

\section{Polynomials of hypergeometric type}

It is well-known \cite{NSU} that for $\lambda =\lambda _l$, where $l\in \mathbb{N}$ and
\begin{equation}
\lambda _l=-\frac{\sigma ''}{2}l(l-1)-\tau 'l
=\left\{ 
\begin{array}{lcl}
-\alpha l & {\rm if} & \sigma ''=0\\[2mm]
l(l-\alpha -1) & {\rm if } & \sigma ''=-2\\[2mm]
l(1-\alpha -l) & {\rm if } & \sigma ''=2
\end{array} \right.
\end{equation}
the equation (\ref{hypeq}) admits a polynomial solution 
$\Phi _l=\Phi _l^{(\alpha ,\beta )}$ of at most $l$ degree
\begin{equation} \label{eq3}
\sigma (s) \Phi _l ''+\tau (s) \Phi _l '+\lambda _l\Phi _l=0.
\end{equation}
{\bf Theorem 1}. {\it 
The function $\Phi _l(s)\sqrt{\varrho (s)}$ is square integrable on $(a,b)$ and
\[ 0=\lambda _0<\lambda _1<\lambda _2<\, \dots \, <\lambda _l \]
for any $l<\Lambda $, where} 
\begin{equation}
\Lambda =\left\{ \begin{array}{lcl}
\infty & if & \sigma ''\in \{ 0,-2\}\\[1mm]
\frac{1-\alpha }{2} & \mbox{}\quad  if \quad \mbox{} & 
\sigma ''=2 .\\[2mm]
\end{array}\right.
\end{equation} 
{\bf Proof.}
The convergence of the integral $ \int_a^b|\Phi _l(s)|^2\varrho (s)ds $
follows from the restrictions imposed to $\alpha $ and $\beta $ (see table 1) and from the relation
\begin{equation}\label{gamma1} \fl
\left\{ \gamma \ | \ 
\lim_{s\rightarrow a}\sigma (s)\varrho (s)s^\gamma\!=\!0\ \ {\rm and}\ \ 
\lim_{s\rightarrow b}\sigma (s)\varrho (s)s^\gamma \!=\!0 \right\}
\!=\!\left\{ \begin{array}{lll}
[0,\infty ) & {\rm if} & \sigma ''\!\in \!\{ 0,\, -2\}\\[1mm]
[0,-\alpha ) & {\rm if} & \sigma ''=2 . 
\end{array} \right.  
\end{equation}
$\opensquare $\\[5mm]
{\bf Theorem 2. }
{\it The system of polynomials
$\{\Phi _l\ |\ l<\Lambda \}$ is  orthogonal 
with weight function $\varrho (s)$ in $(a,b)$, and
$\Phi _l$ is a polynomial of degree $l$ for any $l<\Lambda $.}\\[5mm]
{\bf Proof.} 
Let $l,\ k\in \mathbb{N}$ with $0\leq l<k<\Lambda $. From the relations 
\[ [\sigma (s) \varrho (s)\Phi '_l]'+\lambda _l\varrho (s)\Phi _l =0\quad {\rm and}\quad  
[\sigma (s) \varrho (s)\Phi '_k]'+\lambda _k\varrho (s)\Phi _k =0 \]
we get
\[ (\lambda _l-\lambda _k)\Phi _l(s)\Phi _k(s)\varrho (s)
=\frac{d}{ds}\{ \sigma (s)\varrho (s) 
[\Phi _l(s)\Phi '_k(s)-\Phi _k(s)\Phi '_l(s)]\} .\]
Since the Wronskian $W[\Phi _l(s),\Phi _k(s)]
=\Phi _l(s)\Phi '_k(s)-\Phi _k(s)\Phi '_l(s)$ is a polynomial of 
at most $l+k-1$ degree, from (\ref{gamma1}) it follows that
\[ (\lambda _l-\lambda _k)\int_a^b\Phi _l(s)\Phi _k(s)\varrho (s)ds
=\sigma (s)\varrho (s)W[\Phi _l(s),\Phi _k(s)]|_a^b=0.\]
Each $\Phi _l$ is a polynomial of at most $l$ degree and
the polynomials $\{ \Phi _l\ |\ 0\leq l<\Lambda \}$ are 
linearly independent. This is possible only if $\Phi _l$ is a
polynomial of degree $l$ for any $l<\Lambda $.\qquad $\opensquare $\\[5mm]
{\bf Theorem 3.}
{\it Up to a multiplicative constant
\begin{equation}\label{classical}
  \Phi _l^{(\alpha ,\beta )}(s)\!=\!\left\{ \begin{array}{lcl}
{\bf H}_l\left(\sqrt{\frac{-\alpha }{2}}\, s-\frac{\beta }{\sqrt{-2\alpha }}\right)  
& {\mbox{}\quad if \quad \mbox{}} & \sigma (s)=1\\[2mm]
{\bf L}_l^{\beta -1}(-\alpha s)  & if & \sigma (s)=s\\[2mm]
{\bf P}_l^{(-(\alpha +\beta )/2-1,\ (-\alpha +\beta )/2-1)}(s)  & if & \sigma (s)\!=\!1-s^2\\[2mm]
{\bf P}_l^{((\alpha -\beta )/2-1,\ (\alpha +\beta )/2-1)}(-s)  & if & \sigma (s)\!=\!s^2-1\\[2mm]
\left(\frac{s}{\beta }\right)^l {\bf L}_l^{1-\alpha -2l}\left(\frac{\beta }{s}\right) 
& if & \sigma (s)=s^2\\[2mm]
{\rm i}^l {\bf P}_l^{((\alpha +{\rm i}\beta )/2-1,\ (\alpha -{\rm i}\beta )/2-1)}({\rm i}s) 
& if & \sigma (s)\!=\!s^2+1
\end{array} \right.
\end{equation}
where ${\bf H}_n$, ${\bf L}_n^p $ and ${\bf P}_n^{(p,q)}$ are the Hermite,
Laguerre and Jacobi polynomials, respectively.}\\[5mm]
{\bf Proof.}
In the case $\sigma (s)=s^2$ the function $\Phi _l^{(\alpha ,\beta )}(s)$
satisfies the equation
\[ s^2y''+(\alpha s+\beta )y'+[-l(l-1)-\alpha l]y=0. \]
If we denote $t=\beta /s$ then the polynomial $u(t)=t^ly(\beta /t)$ satisfies the equation
\[ tu''+(-\alpha +2-2l-t)u'+lu=0 \]
that is, the equation whose polynomial solution is ${\bf L}_l^{1-\alpha -2l}(s)$ (up to
a multiplicative constant).
In a similar way one can analyse the other cases.\qquad $\opensquare $
\section{Special functions of hypergeometric type}
Let $l\in \mathbb{N}$, $l<\Lambda $, and let $m\in \{ 0,1,...,l\}$.
If we differentiate (\ref{eq3}) $m$ times then we get  
\begin{equation}\label{varphi}
 \sigma (s)\frac{{\rm d}^{m+2}}{{\rm d}s^{m+2}}\Phi _l
+[\tau (s)+m\sigma '(s)]\frac{{\rm d}^{m+1}}{{\rm d}s^{m+1}}\Phi _l
+(\lambda _l-\lambda _m) \frac{{\rm d}^m}{{\rm d}s^m}\Phi _l=0.
\end{equation}
The equation obtained by multiplying this relation by $\sqrt{\sigma ^m(s)}$ can be written as
\begin{equation}\label{Hm}
H_m \Phi _{l,m}=\lambda _l\Phi _{l,m}
\end{equation}
where $H_m$ is the differential operator
\begin{equation} \label{defHm} 
\fl H_m =-\sigma (s) \frac{d^2}{ds^2}-\tau (s) \frac{d}{ds}
+\frac{m(m-2)}{4}\frac{(\sigma '(s))^2}{\sigma (s)}   
 + \frac{m\tau (s)}{2}\frac{\sigma '(s)}{\sigma (s)}
-\frac{1}{2}m(m-2)\sigma ''(s)-m\tau '(s) . 
\end{equation}
and the functions 
\begin{equation}\label{def}
\Phi _{l,m}(s)=\kappa ^m(s)\frac{{\rm d}^m}{{\rm d}s^m}\Phi _l(s) 
\end{equation}
defined by using 
\begin{equation}
\kappa (s)=\sqrt{\sigma (s)}
\end{equation}  
are called the {\em associated special functions}.\\[5mm]
{\bf Theorem 4.}
{\it If \ $0\leq m\leq l<\Lambda $ \ then \ $\Phi _{l,m}(s)\sqrt{\varrho (s)}$
is square integrable on $(a,b)$.}\\[5mm]
{\bf Proof.}
The convergence of the integral $ \int_a^b|\Phi _{l,m}(s)|^2\varrho (s)ds $
follows from the restrictions imposed to $\alpha $ and $\beta $ (see table 1) 
and from the relation (\ref{gamma1}).\qquad $\opensquare $\\[5mm]
By differentiating (\ref{eq3}) $m-1$ times we obtain
\[   \begin{array}{r}
\sigma (s)\frac{{\rm d}^{m+1}}{{\rm d}s^{m+1}}\Phi _l(s)
+(m-1)\sigma '(s)\frac{{\rm d}^m}{{\rm d}s^m}\Phi _l(s)+
\frac{(m-1)(m-2)}{2}\sigma ''(s)\frac{{\rm d}^{m-1}}{{\rm d}s^{m-1}}\Phi _l(s)\\[2mm]
+\tau (s)\frac{{\rm d}^m}{{\rm d}s^m}\Phi _l
+(m-1)\tau '(s)\frac{{\rm d}^{m-1}}{{\rm d}s^{m-1}}\Phi _l(s)
+\lambda _l\frac{{\rm d}^{m-1}}{{\rm d}s^{m-1}}\Phi _l(s)=0.\end{array}\]
If we multiply this relation by $\kappa ^{m-1}(s)$ then we get 
the three term recurrence relation
\begin{equation} 
\begin{array}{l}
   \Phi _{l,m+1}(s) \!+\! \left( \frac{\tau (s)}{\kappa (s)}
\!+\!2(m\!-\!1)\kappa '(s)\right)\Phi _{l,m}(s) 
\!+\!(\lambda _l\!-\!\lambda _{m-1}) \Phi _{l,m-1}(s)\!=\!0 \label{rec}
\end{array}
\end{equation}
for $m\in \{ 1,2,...,l-1\}$, and 
\begin{equation}\label{rec1}
\left( \frac{\tau (s)}{\kappa (s)}+
2(l-1)\kappa '(s)\right) \Phi _{l,l}(s)
+(\lambda _l-\lambda _{l-1})\Phi _{l,l-1}(s)=0
\end{equation}
for $m=l$. 
For each $m\in \{ 0,1,...,l-1\}$, 
by differentiating (\ref{def}), we obtain
\[ \frac{{\rm d}}{{\rm d}s}\Phi _{l,m}(s)
=m\kappa ^{m-1}(s)\kappa '(s)\frac{{\rm d}^m}{{\rm d}s^m}\Phi _l
+\kappa ^m(s)\frac{{\rm d}^{m+1}}{{\rm d}s^{m+1}}\Phi _l(s) \]
that is, the relation 
\[ \kappa (s)\frac{{\rm d}}{{\rm d}s}\Phi _{l,m}(s)=m\kappa '(s)\Phi _{l,m}(s)+\Phi_{l,m+1}(s) \]
which can be written as
\begin{equation}
\left(\kappa (s)\frac{d}{ds}-
m\kappa '(s)\right) \Phi _{l,m}(s)=\Phi _{l,m+1}(s).\label{raising}
\end{equation}
If $m\in \{ 1,2,...,l-1\}$ then by substituting (\ref{raising}) into
(\ref{rec}) we get 
\[   \left( \kappa (s)\frac{d}{ds}+\frac{\tau (s)}{\kappa (s)}
+(m-2)\kappa '(s)\right)\Phi _{l,m}(s) 
+(\lambda _l-\lambda _{m-1}) \Phi _{l,m-1}(s)=0 \]
that is,
\begin{equation}
  \left(-\kappa (s)\frac{d}{ds}-
   \frac{\tau (s)}{\kappa (s)}-(m-1)\kappa '(s)\right)
\Phi _{l,m+1}(s)=(\lambda _l-\lambda _m)\Phi _{l,m}(s).\label{lowering}
\end{equation}
for all $m\in \{ 0,1,...,l-2\}$. From (\ref{rec1}) it follows that this 
relation is also satisfied for $m=l-1$.

The relations (\ref{raising})  and (\ref{lowering}) suggest us to consider 
for $m+1<\Lambda $ the operators
\begin{equation}\begin{array}{l}
  A_m=\kappa (s)\frac{d}{ds}-m\kappa '(s)\\[2mm]
A_m^+=-\kappa (s)\frac{d}{ds}-\frac{\tau (s)}{\kappa (s)}-(m-1)\kappa '(s).
\end{array}
\end{equation}
satisfying the relations
\begin{equation}\label{AmAm+}
\begin{array}{l}
A_m\Phi _{l,m}=\left\{ \begin{array}{lll}
0 & {\rm for} & l=m\\
\Phi _{l,m+1} & {\rm for} & m<l<\Lambda 
\end{array} \right. \\[5mm]
A_m^+\Phi _{l,m+1}\!=\!(\lambda _l\!-\!\lambda _m)\Phi _{l,m}\ \  
{\rm for}\ \ 0\leq m<l< \Lambda .
\end{array}
\end{equation}
and 
\begin{equation}\label{philm}
\Phi _{l,m}(s)=\left\{ \begin{array}{lll}
\kappa ^l(s) & {\rm for} & m=l\\
\frac{A_m^+ }{\lambda _l-\lambda _m}
\frac{A_{m+1}^+ }{\lambda _l-\lambda _{m+1}}...
\frac{A_{l-1}^+ }{\lambda _l-\lambda _{l-1}}\kappa ^l(s) \quad & {\rm for} \quad & 0<m<l<\Lambda .
\end{array} \right.
\end{equation}
\mbox{}\\[3mm]
{\bf Theorem 5.}
{\it The operators $H_m$ are shape invariant
\begin{equation}
H_m-\lambda _m=A_m^+A_m\qquad H_{m+1}-\lambda _m=A_mA_m^+  
\end{equation} 
and satisfy the intertwining relations}
\begin{equation}
\quad H_mA_m^+=A_m^+H_{m+1}\qquad A_mH_m=H_{m+1}A_m.
\end{equation}
{\bf Proof.}
Direct computation.\qquad $\opensquare $\\[5mm]
{\bf Lemma 1.}
{\it If the functions $\varphi ,\psi :(a,b)\longrightarrow \mathbb{R}$ are derivable and if 
\begin{equation}
 \lim_{s\rightarrow a}\kappa (s)\varrho (s)\, \varphi (s)\, \psi (s)=
\lim_{s\rightarrow b}\kappa (s)\varrho (s)\, \varphi (s)\, \psi (s)=0
\end{equation}
then 
\begin{equation}
\langle A_m\varphi ,\psi \rangle =\langle  \varphi ,A_m^+\psi \rangle \qquad and \qquad 
\langle A_m^+\varphi ,\psi \rangle =\langle  \varphi ,A_m\psi \rangle.
\end{equation}
If the functions $\varphi ,\psi :(a,b)\longrightarrow \mathbb{R}$ are twice derivable and if 
\begin{equation}
\begin{array}{l}
\lim_{s\rightarrow a}\kappa (s)\varrho (s)\, (A_m \varphi )(s)\, \psi (s)=
\lim_{s\rightarrow b}\kappa (s)\varrho (s)\, (A_m \varphi )(s)\, \psi (s)=0\\
\lim_{s\rightarrow a}\kappa (s)\varrho (s)\, \varphi (s) (A_m \psi ) (s)=
\lim_{s\rightarrow b}\kappa (s)\varrho (s)\, \varphi (s) (A_m \psi ) (s)=0
\end{array}
\end{equation}
then} 
\begin{equation}
\langle H_m\varphi ,\psi \rangle =\langle  \varphi ,H_m\psi \rangle . 
\end{equation}
{\bf Proof.}
Integrating by parts we get 
\[ \begin{array}{r}
 \langle A_m \varphi ,\psi \rangle 
\!=\!\!\int_a^b\left[\kappa (s)\frac{d}{ds}-m\kappa '(s)\right] 
\varphi (s)\,\psi (s)\, \varrho (s)ds 
=\kappa (s)\varrho(s)\varphi (s) \psi(s)|_a^b \\[2mm]
+\int_a^b\!\varphi (s)\left[-\kappa (s)\frac{d}{ds}\!-\!\frac{\tau (s)}{\kappa (s)}-\!(m\!-\!1)\kappa '(s)
\right] \psi (s)\varrho (s)ds
= \langle \varphi ,A_m^+\psi \rangle   
\end{array} \]
and
$\langle H_m\varphi ,\psi \rangle \!=\!\langle (A_m^+A_m\!+\!\lambda _m)\varphi ,\psi \rangle \!=\!
\langle A_m\varphi ,A_m\psi \rangle \!+\!\langle \varphi ,\lambda _m\psi \rangle 
\!=\!\langle  \varphi ,H_m\psi \rangle . $\qquad $\opensquare $\\[5mm]
{\bf Theorem 6.}
{\it For each $m<\Lambda $, the functions
$\Phi _{l,m}$ with $m\leq l<\Lambda $ 
are orthogonal with weight function $\varrho (s)$ in $(a,b)$.}\\[5mm]
{\bf Proof.}
By using (\ref{gamma1}) we get
\[ \lim_{s\rightarrow a,b}\kappa (s)\varrho (s)(A_m \Phi _{l,m})(s) \Phi _{k,m}(s)\!=\!
\lim_{s\rightarrow a,b}\kappa (s)\varrho (s)\Phi _{l,m+1}(s) \Phi _{k,m}(s)=0\]
for $l,k\!\in \!\mathbb{N}$ with $m\!\leq \!l\!<\!\Lambda $ and $m\!\leq \!k\!<\!\Lambda $ . 
Therefore, we can use lemma 1 and obtain 
\[ (\lambda _l-\lambda _k)\langle \Phi _{l,m},\Phi _{k,m}\rangle =
\langle H_m\Phi _{l,m},\Phi _{k,m}\rangle -\langle \Phi _{l,m},H_m\Phi _{k,m}\rangle =0.\qquad \opensquare \]
{\bf Theorem 7.}
{\it If \ $0\!\leq m<l\!<\! \Lambda $ \ then} \
$||\Phi _{l,m+1}||\!=\!\sqrt{\lambda _l\!-\!\lambda _m}\, ||\Phi _{l,m}|| $.\\[5mm]
{\bf Proof.}
Since 
$\lim_{s\rightarrow a}\kappa (s)\varrho (s)\Phi _{l,m}(s) \Phi _{l,m+1}(s)\!=\!
\lim_{s\rightarrow b}\kappa (s)\varrho (s)\Phi _{l,m}(s) \Phi _{l,m+1}(s)=0$ we get
\[ ||\Phi _{l,m+1}||^2\!=\!\langle A_m\Phi _{l,m},\Phi _{l,m+1}\rangle 
 \!=\!  \langle \Phi _{l,m},A_m^+\Phi _{l,m+1}\rangle
\!=\!(\lambda _l\!-\!\lambda _m)||\Phi _{l,m}||^2 .\qquad \opensquare \]
The {\em normalized associated special functions} 
\begin{equation}
 \phi _{l,m}=\Phi _{l,m}/||\Phi _{l,m}||
\end{equation}
satisfy the relations
\begin{equation}
\begin{array}{l}
A_m\ \phi _{l,m}=\left\{ \begin{array}{lll}
0 & {\rm for} & l=m\\
\sqrt{\lambda _l-\lambda _m}\ \phi _{l,m+1} & {\rm for} & m<l<\Lambda 
\end{array} \right.\\[5mm]
A_m^+\ \phi _{l,m+1}=\sqrt{\lambda _l-\lambda _m}\ \phi _{l,m}
\quad {\rm for}\  0\leq m<l<\Lambda \\[3mm]
\phi _{l,m}=
\frac{A_m^+ }{\sqrt{\lambda _l-\lambda _m}}
\frac{A_{m+1}^+ }{\sqrt{\lambda _l-\lambda _{m+1}}}...
\frac{A_{l-1}^+ }{\sqrt{\lambda _l-\lambda _{l-1}}}\phi _{l,l}. 
\end{array}
\end{equation}
\section{Shape invariant operators related to $H_m$}
Some additional shape invariant operators directly related to $H_m$ can be obtained in the 
cases when $\alpha $ and $\beta $ are such that 
there exists $k\in \mathbb{R}$ with $\varrho (s)=\sigma ^k(s)$ (see table 2).

\begin{table}[htbp]
\caption{The cases when $\varrho (s)=\sigma ^k(s)$ }
\begin{center}
\begin{tabular}{cllrl}
\hline
$\sigma (s)$ & \quad $\tau (s) $  & \quad $\varrho (s)$  & $k$\quad \mbox{} &  \quad $(a,b) $\\
\hline \hline
$s$ & \quad $\beta $ & \quad $s^{\beta -1}$ & \quad $\beta -1$& $\quad (0,\infty )$\\ 
$1-s^2$  &\quad  $\alpha s$ &\quad  $(1-s^2)^{-\alpha /2-1}$& \quad $-\frac{\alpha }{2}-1$ & \quad $(-1,1)$\\
$s^2-1$  &\quad  $\alpha s$ &\quad  $(s^2-1)^{\alpha /2-1}$ & \quad $\frac{\alpha }{2}-1$ & \quad $(1,\infty )$\\
$s^2$  & $\quad \alpha s$ & \quad $s^{\alpha -2}$& \quad $\frac{\alpha }{2}-1$ & \quad $(0,\infty )$\\
$s^2+1$  & \quad $\alpha s$ &\quad  $(s^2+1)^{\alpha /2-1}$ &\quad  $\frac{\alpha }{2}-1$ & \quad $(-\infty ,\infty )$\\
\hline
\end{tabular}
\end{center}
\end{table}

From $(\sigma \varrho )'=\tau \varrho $
we get $\tau (s)=(k+1)\sigma '(s)=2(k+1)\kappa (s)\, \kappa '(s)$, and
\begin{equation} \begin{array}{l}
  A_m=\kappa (s)\frac{d}{ds}-m\kappa '(s)\qquad
A_m^+=-\kappa (s)\frac{d}{ds}-(2k+m+1)\kappa '(s).
\end{array} \end{equation}
{\bf Theorem 8.}
{\it If $\alpha $ and $\beta $ are such that $\varrho (s)=\sigma ^k(s)$
then for any $\delta \in \mathbb{R}$ the operators
\begin{equation}
\tilde{A}_m=A_m+\frac{\delta }{2m+2k+1}\qquad \quad
\tilde{A}_m^+=A_m^++\frac{\delta }{2m+2k+1} 
\end{equation}
satisfy the relations
\begin{equation} \label{shapeinvar2} 
\begin{array}{ll}
\tilde{A}_m^+\tilde{A}_m=\tilde{H}_m-\tilde{\lambda }_m\qquad & 
\tilde{A}_m\tilde{H}_m=\tilde{H}_{m+1}\tilde{A}_m\\[2mm]
\tilde{A}_m\tilde{A}_m^+=\tilde{H}_{m+1}-\tilde{\lambda }_m \qquad &
\tilde{H}_m\tilde{A}_m^+=\tilde{A}_m^+\tilde{H}_{m+1}
\end{array}
\end{equation}
where  
\begin{equation}\label{Lma}
\tilde{H}_m={H}_m-\delta \frac{d\kappa }{ds}\qquad \qquad 
\tilde{\lambda }_m=\lambda _m-\frac{\delta ^2}{(2m+2k+1)^2}
\end{equation}
for any $m\in \mathbb{R}$ with $2m+2k+1\not=0.$}\\[5mm]
{\bf Proof.}
Since $A_m^+A_m={H}_m-\lambda _m$ and $A_mA_m^+={H}_{m+1}-\lambda _m$ we obtain
\[ \begin{array}{l}
(A_m^++\varepsilon )(A_m+\varepsilon )={H}_m-\lambda _m-\varepsilon (2m+2k+1)\kappa '(s)+\varepsilon ^2 \\
(A_m+\varepsilon )(A_m^++\varepsilon )={H}_{m+1}-\lambda _m-\varepsilon (2m+2k+1)\kappa '(s)+\varepsilon ^2 
\end{array} \]
for any constant $\varepsilon $. If we choose $\varepsilon =\delta /(2m+2k+1)$ then we get (\ref{shapeinvar2})
\[ \begin{array}{l}
\tilde{H}_m\tilde{A}_m^+=(\tilde{A}_m^+\tilde{A}_m+\tilde{\lambda }_m)\tilde{A}_m^+=
\tilde{A}_m^+(\tilde{A}_m\tilde{A}_m^++\tilde{\lambda }_m)=\tilde{A}_m^+\tilde{H}_{m+1}\\
\tilde{A}_m\tilde{H}_m=\tilde{A}_m(\tilde{A}_m^+\tilde{A}_m+\tilde{\lambda }_m)=
(\tilde{A}_m\tilde{A}_m^++\tilde{\lambda }_m)\tilde{A}_m=\tilde{H}_{m+1}\tilde{A}_m.\qquad \opensquare  
\end{array} \]
The mapping $m\mapsto \tilde \lambda _m$ is an increasing function on the set 
$\{ m\ | \ \frac{{\rm d}}{{\rm d}m}\tilde \lambda _m>0\}$, where
\begin{equation} 
\frac{{\rm d}}{{\rm d}m}\tilde \lambda _m
\!=\!\left\{ 
\begin{array}{rcl}
4\delta ^2(2m\!+\!2\beta \!-\!1)^{-3} & {\rm if} & \sigma (s)=s\\[3mm]
(2m\!-\!\alpha -1)\left[ 1\!+\!4\delta ^2(2m\!-\!\alpha \!-\!1)^{-4}\right]
& {\rm if } & \sigma (s)\!=\!1\!-\!s^2\\[3mm]
(2m\!+\!\alpha \!-\!1)\left[ -1\!+\!4\delta ^2(2m\!+\!\alpha \!-\!1)^{-4}\right]
 & {\rm if } & \sigma ''=2
\end{array} \right.
\end{equation}
and, up to a multiplicative constant, the solution $\tilde \Phi _{m,m}$ 
of the equation $\tilde A_m\tilde \Phi _{m,m}\!=\!0$ is
\begin{equation} 
\tilde \Phi _{m,m}(s)\!=\!\left\{ 
\begin{array}{rcl}
(\sqrt{s})^m\, {\rm e}^{ -\frac{2\delta \, \sqrt{s}}{2m+2\beta -1}} & {\rm if} & \sigma (s)=s\\[3mm]
(\sqrt{1\!-\!s^2})^m\, {\rm e}^{-\frac{\delta \, {\rm arcsin}s}{2m-\alpha -1} }
& {\rm if } & \sigma (s)\!=\!1\!-\!s^2\\[3mm]
(\sqrt{s^2\!-\!1})^m\, (s\!+\!\sqrt{s^2\!-\!1})^{-\frac{\delta }{2m+\alpha -1} }
& {\rm if } & \sigma (s)\!=\!s^2\!-\!1\\[3mm]
s^{m-\frac{\delta }{2m+\alpha -1}}
 & {\rm if } & \sigma (s)=s^2\\[3mm]
(\sqrt{s^2\!+\!1})^m\, (s\!+\!\sqrt{s^2\!+\!1})^{-\frac{\delta }{2m+\alpha -1} }
& {\rm if } & \sigma (s)\!=s^2\!+\!1.
\end{array} \right.
\end{equation}
The set $\mathcal{M}=\{ \ m\ |\ \frac{{\rm d}}{{\rm d}m}\tilde \lambda _m>0 \ 
{\rm and}\ \int_a^b\tilde \Phi ^2_{m,m}(s)\varrho (s) ds<\infty \  \}$
of all the values of $m$ for which $\frac{{\rm d}}{{\rm d}m}\tilde \lambda _m>0$ and
$\tilde \Phi _{m,m}\sqrt{\varrho }$ is square integrable on $(a,b)$  is presented in table 3. 

\begin{table}[htbp]
\caption{The set $\mathcal{M}$ of all the values of $m$ for which $\frac{{\rm d}}{{\rm d}m}\tilde \lambda _m>0$ and
$\tilde \Phi _{m,m}\sqrt{\varrho }$ is square integrable}
\begin{center}
\begin{tabular}{cll}
\hline 
$\sigma (s)$ & \quad $\tau (s) $\quad \mbox{}  & \ \mbox{}\qquad \qquad $\mathcal{M}$\\
\hline \hline 
$s$ & \quad $\beta $ & $\left\{ \begin{array}{cll}
\emptyset & {\rm for} & \delta \leq 0\\
(-\beta \!+\!\frac{1}{2}, \infty ) & {\rm for} & \delta >0 
\end{array}\right. $\\     
$1-s^2$  &\quad  $\alpha s$ & $\left(\frac{1+\alpha }{2}, \infty \right)$ \ for any \ $\delta \in \mathbb{R}$ \\
$s^2-1$  &\quad  $\alpha s$ & $\left\{ \begin{array}{cll}
\emptyset & {\rm for} & \delta \leq -\frac{1}{2}\\[1mm]
\left( -\frac{\alpha }{2}, \frac{1-\alpha }{2}\!-\!\sqrt{-\frac{\delta }{2}}\right) 
& {\rm for } & \delta \!\in \!(-\frac{1}{2}, 0]\\[1mm]
\left( -\frac{\alpha }{2}, \frac{1-\alpha }{2}\!-\!\sqrt{-\frac{\delta }{2}}\right) 
\cup  \left( \frac{1-\alpha }{2}, \frac{1-\alpha }{2}\!+\!\sqrt{\frac{\delta }{2}}\right) 
& {\rm for } & \delta \in (0,\frac{1}{2})\\[1mm]
\left( \frac{1-\alpha }{2}, \frac{1-\alpha }{2}\!+\!\sqrt{\frac{\delta }{2}}\right) 
& {\rm for } & \delta \in [\frac{1}{2}, \infty )
\end{array}\right. $\\    
$s^2$  & $\quad \alpha s$ &  $\emptyset $ \ for any \ $\delta \in \mathbb{R}$ \\[1mm]
$s^2+1$  & \quad $\alpha s$ & $\left(-\infty , \frac{1-\alpha }{2}\!-\!\sqrt{\frac{|\delta |}{2}}\right)$ 
\ for any \ $\delta \in \mathbb{R}$  \\
\hline 
\end{tabular}
\end{center}
\end{table}

\noindent {\bf Lemma 2.}
{\it If $l\in \mathcal{M}$ and $n\in \mathbb{N}$ are such that $\{ l-n,l-n+1,...,l\}\subset \mathcal{M}$
then for each $m\in \{ l-n,l-n+1,...,l-1\}$ the function 
\begin{equation} \label{tildephi}
\tilde{\Phi }_{l,m}=
\frac{\tilde{A}_m^+ }{\tilde{\lambda }_l-\tilde{\lambda }_m}\
\frac{\tilde{A}_{m+1}^+ }{\tilde{\lambda }_l-\tilde{\lambda }_{m+1}}\ ...\
\frac{\tilde{A}_{l-2}^+ }{\tilde{\lambda }_l-\tilde{\lambda }_{l-2}}\
\frac{\tilde{A}_{l-1}^+ }{\tilde{\lambda }_l-\tilde{\lambda }_{l-1}}\, \tilde{\Phi }_{l,l}
\end{equation}
has the form
\begin{equation} \label{phiform}
\tilde{\Phi }_{l,m}\!=\!\left\{
\begin{array}{lcl}
\sum_{j=0}^{l-m}c_j(\sqrt{s})^{l-j}\,
 {\rm e}^{ -\frac{2\delta \, \sqrt{s}}{2l+2\beta -1}} &  if & \sigma (s)=s\\[3mm]
\sum_{j=0}^{l-m}c_js^j(\sqrt{1\!-\!s^2})^{l-j}\,
{\rm e}^{-\frac{\delta \, {\rm arcsin}s}{2l\!-\!\alpha \!-\!1} } & if & \sigma (s)\!=\!1\!-\!s^2\\[3mm]
\sum_{j=0}^{l-m}c_js^j(\sqrt{s^2\!-\!1})^{l-j}\,
(s\!+\!\sqrt{s^2\!-\!1})^{-\frac{\delta }{2l\!+\!\alpha \!-\!1} }& if & \sigma (s)\!=\!s^2\!-\!1\\[3mm]
\sum_{j=0}^{l-m}c_js^j(\sqrt{s^2\!+\!1})^{l-j}\,
(s\!+\!\sqrt{s^2\!+\!1})^{-\frac{\delta }{2l\!+\!\alpha \!-\!1} }& if & \sigma (s)\!=s^2\!+\!1
\end{array} \right.
\end{equation}
where $c_j$ are real constants.}\\[5mm]
{\bf Proof.}
The statement follows from the relations
\[  \fl \begin{array}{rl}
\frac{d}{ds}\sqrt{s}=\frac{1}{2\sqrt{s}} \qquad \mbox{}& 
\sqrt{s}\frac{d}{ds}{\rm e}^{-\frac{2\delta \sqrt{s}}{2l+2\beta -1}}
=-\frac{\delta }{2l+2\beta -1}{\rm e}^{-\frac{2\delta \sqrt{s}}{2l+2\beta -1}}\\[2mm]
\frac{d}{ds}\sqrt{1\!-\!s^2}\!=\!\frac{-s}{\sqrt{1\!-\!s^2}} \qquad \mbox{}& 
\sqrt{1-s^2}\frac{d}{ds}{\rm e}^{-\frac{\delta \arcsin {s}}{2l-\alpha -1}}
=-\frac{\delta }{2l-\alpha -1}{\rm e}^{-\frac{\delta \arcsin {s}}{2l-\alpha -1}}\\[2mm]
\frac{d}{ds}\sqrt{s^2\!\mp \!1}\!=\!\frac{s}{\sqrt{s^2\!\mp \!1}} \qquad \mbox{}& 
\sqrt{1\!-\!s^2}\frac{d}{ds}(s\!+\!\sqrt{s^2\mp 1})^{-\frac{\delta }{2l\!+\alpha -\!1}}
\!=\!-\frac{\delta }{2l\!+\alpha -\!1}(s\!+\!\sqrt{s^2\mp 1})^{-\frac{\delta }{2l\!+\alpha -\!1}}
\end{array} \]
$\opensquare $\\[5mm] 
{\bf Theorem 9.}
{\it If $l\!\in \!\mathcal{M}$ and $n\!\in \!\mathbb{N}$ are such that $ \{ l\!-n,l\!-n\!+\!1,...,l\}\!\subset \!\mathcal{M} $
then for each $m\in \{ l-n,l-n+1,...,l\}$ the function $\tilde \Phi _{l,m}\sqrt{\varrho }$ 
is square integrable and}
\begin{equation}\label{schrod2}
\tilde{H}_m\tilde{\Phi }_{l,m}=\tilde{\lambda }_l\tilde{\Phi }_{l,m}
\end{equation} 
\begin{equation}\begin{array}{l}
\tilde{A}_m\tilde{\Phi }_{l,m}=\left\{ \begin{array}{lll}
0 & for & m=l\\
\tilde{\Phi }_{l,m+1} & for & m<l
\end{array} \right. \\
\tilde{A}_m^+\tilde{\Phi }_{l,m+1}=(\tilde{\lambda }_l-\tilde{\lambda }_m)\tilde{\Phi }_{l,m} \qquad  for\ \ m<l.
\end{array}
\end{equation}
{\bf Proof.}
The square integrability follows from (\ref{phiform}). 
The definition (\ref{tildephi}) of $\tilde{\Phi }_{l,m}$ can be re-written as
\begin{equation} \label{rectildephi}
\tilde{\Phi }_{l,m}=
\frac{\tilde{A}_m^+ }{\tilde{\lambda }_l-\tilde{\lambda }_m}\, \tilde{\Phi }_{l,m+1}
\end{equation}
and $\tilde{H}_l\tilde{\Phi }_{l,l}
\!=\!(\tilde{A}_l^+\tilde{A}_l+\tilde{\lambda }_l)\tilde{\Phi }_{l,l}\!=\!\tilde{\lambda }_l\tilde{\Phi }_{l,l}$.
The relation $\tilde{H}_m\tilde{\Phi }_{l,m}\!=\!\tilde{\lambda }_l\tilde{\Phi }_{l,m}$ follows by recurrence
\[  \fl  
\tilde{H}_{m+1}\tilde{\Phi }_{l,m+1}\!=\!\tilde{\lambda }_l\tilde{\Phi }_{l,m+1}\quad \Longrightarrow \quad 
\tilde{H}_m\tilde{\Phi }_{l,m}
\!=\!\frac{\tilde{H}_m\tilde{A}_m^+ }{\tilde{\lambda }_l-\tilde{\lambda }_m}\, \tilde{\Phi }_{l,m+1}
\!=\!\frac{\tilde{A}_m^+ \tilde{H}_{m+1} }{\tilde{\lambda }_l-\tilde{\lambda }_m}\, \tilde{\Phi }_{l,m+1}
\!=\!\tilde{\lambda }_l\tilde{\Phi }_{l,m}.
\]
From the relation (\ref{rectildephi}) we get
\[
\tilde{A}_m\tilde{\Phi }_{l,m}
=\frac{\tilde{A}_m\tilde{A}_m^+ }{\tilde{\lambda }_l-\tilde{\lambda }_m}\, \tilde{\Phi }_{l,m+1}
=\frac{\tilde{H}_{m+1}-\tilde{\lambda }_m}{\tilde{\lambda }_l-\tilde{\lambda }_m}\, \tilde{\Phi }_{l,m+1}
=\tilde{\Phi }_{l,m+1}.\qquad \opensquare  \]
{\bf Theorem 10.}
{\it If $l\in \mathcal{M}$ and $n\in \mathbb{N}$ are such that $\{ l\!-\!n\!-\!1,l\!-\!n,...,l\}\!\subset \!\mathcal{M}$ 
then the functions $\tilde \Phi _{m,m}$, $\tilde \Phi _{m+1,m}$, ... , $\tilde \Phi _{l,m}$, where \ $m\!=\!l\!-\!n$, are orthogonal.}\\[5mm]
{\bf Proof.}
By using (\ref{phiform}) and lemma 1 we get 
\[ (\tilde \lambda _{n'}-\tilde \lambda _{n''}) \langle \tilde \Phi _{n',m},\tilde \Phi _{n'',m}\rangle=
\langle \tilde H_m\Phi _{n',m},\tilde \Phi _{n'',m}\rangle -\langle \tilde \Phi _{n',m},H_m\tilde \Phi _{n'',m}\rangle =0.\]
for any $n',\, n''\in \{ m,m+1, ..., l\}$.\qquad $\opensquare $\\[5mm]
{\bf Theorem 11.}
{\it If $l\in \mathcal{M}$ and $n\in \mathbb{N}$ are such that 
\[ \{ l-n,l-n+1,...,l\}\subset \mathcal{M} \]
then for each $m\in \{ l-n,l-n+1,...,l-1\}$ we have
\begin{equation} 
\langle \tilde A_m\tilde \Phi _{l,m},\tilde \Phi _{l,m+1} \rangle 
=\langle \tilde \Phi _{l,m} ,\tilde A_m^+\tilde \Phi _{l,m+1} \rangle 
\end{equation} 
and}
\begin{equation} 
\quad ||\tilde \Phi _{l,m+1}||=\sqrt{\tilde \lambda _l-\tilde \lambda _m}\, ||\tilde \Phi _{l,m}|| .
\end{equation} 
{\bf Proof.}
We have 
$\lim_{s\rightarrow a}\kappa ^{2k+1}(s)\tilde \Phi _{l,m}(s)\tilde \Phi _{l,m+1}(s)\!=\!
\lim_{s\rightarrow b}\kappa ^{2k+1}(s)\tilde \Phi _{l,m}(s) \tilde \Phi _{l,m+1}(s)\!=\!0$.
In view of lemma 1 we get
\[  \fl  ||\tilde \Phi _{l,m+1}||^2
\!=\!\langle \tilde A_m\tilde \Phi _{l,m},\tilde \Phi _{l,m+1}\rangle 
 \!=\!  \langle \tilde \Phi _{l,m},\tilde A_m^+\tilde \Phi _{l,m+1}\rangle
\!=\!(\tilde \lambda _l\!-\!\tilde \lambda _m)||\tilde \Phi _{l,m}||^2 . \qquad \opensquare \]
The {\em normalized associated special functions} 
\begin{equation}
 \tilde \phi _{l,m}=\tilde \Phi _{l,m}/||\tilde \Phi _{l,m}||
\end{equation}
satisfy the relations
\begin{equation}
\begin{array}{l}
\tilde A_m\ \tilde \phi _{l,m}=\left\{ \begin{array}{lll}
0 & {\rm for} & l=m\\
\sqrt{\tilde \lambda _l-\tilde \lambda _m}\ \tilde \phi _{l,m+1} & {\rm for} & m<l 
\end{array} \right.\\[5mm]
\tilde A_m^+\ \tilde \phi _{l,m+1}=\sqrt{\tilde \lambda _l-\tilde \lambda _m}\ \tilde \phi _{l,m}
\quad {\rm for}\  0\leq m<l \\[3mm]
\tilde \phi _{l,m}=
\frac{\tilde A_m^+ }{\sqrt{\tilde \lambda _l-\tilde \lambda _m}}
\frac{\tilde A_{m+1}^+ }{\sqrt{\tilde \lambda _l-\tilde \lambda _{m+1}}}...
\frac{\tilde A_{l-1}^+ }{\sqrt{\tilde \lambda _l-\tilde \lambda _{l-1}}}\tilde \phi _{l,l}. 
\end{array}
\end{equation}
\section{Gazeau-Klauder coherent states}
Let $m<\Lambda $ be a fixed natural number, and let 
\begin{equation}
\Lambda _m=\Lambda -m=\left\{ \begin{array}{lcl}
\infty & {\rm if} & \sigma ''\in \{ 0,-2\}\\[2mm]
\frac{1-\alpha }{2}-m & {\mbox{}\quad \rm if \quad \mbox{}} & 
\sigma ''=2 .\\[3mm]
\end{array}\right.
\end{equation}
The functions  
\begin{equation}
|n\rangle =\phi _{m+n,m}\qquad {\rm with}\quad 0\leq n<\Lambda _m
\end{equation}
form an orthonormal system, and satisfy the relation
\begin{equation}
(H_m-\lambda _m)|n\rangle =e_n|n\rangle 
\end{equation}
where
\begin{equation} 
e_n=\lambda _{m+n}-\lambda _m
=\left\{ 
\begin{array}{lcl}
-\alpha n & {\rm if} & \sigma ''=0\\[2mm]
n(n+2m-\alpha -1) & {\rm if } & \sigma ''=-2\\[2mm]
n(1-\alpha -2m-n) & {\rm if } & \sigma ''=2.
\end{array} \right.
\end{equation}
One can remark that
\[ 0=e_0<e_1<e_2<\, \dots \, <e_n\qquad {\rm for\ any}\quad n<\Lambda _m .\]
By following the method presented in \cite{GK} we define the Gazeau-Klauder coherent states
\[ |J,\gamma \rangle =N(J)^{-1}\sum_{n<\Lambda _m}\frac{J^{n/2}}{\sqrt{\rho _n}}{\rm e}^{-{\rm i}e_n\gamma }|n\rangle \]
described by the real two-parameter $(J,\gamma )\in [0,\infty )\times \mathbb{R}$, where $N(J)$ is a normalizing constant and 
\begin{equation}
\rho _n=\left\{ \begin{array}{lll}
1 & {\rm if} & n=0\\
e_1e_2\dots e_n & {\rm if} & n>0
\end{array}\right.
=\left\{ 
\begin{array}{lcl}
(-\alpha )^nn! & {\rm if} & \sigma ''=0\\[2mm]
\frac{n!\, \Gamma(n+2m-\alpha )}{\Gamma(2m-\alpha )} & {\rm if } & \sigma ''=-2\\[2mm]
\frac{n!\, \Gamma(1-\alpha -2m)}{\Gamma(1-\alpha -2m-n)}  & {\rm if } & \sigma ''=2.
\end{array} \right.
\end{equation}
and 
\begin{equation}
N(J)=\sqrt{\sum_{n<\Lambda _m}\frac{J^n}{\rho_n}}\ . 
\end{equation}
{\bf Theorem 12.}
{\it There exists a real function $k(J)$ such that
\begin{equation}
\int_0^\infty k(J)\, {\rm d}J \int_{-\infty }^\infty |J,\gamma \rangle \langle J,\gamma |\, {\rm d}\nu (\gamma )
=\sum_{n<\Lambda _m}|n\rangle \langle n|=\mathbb{I} 
\end{equation}
with ${\rm d}\nu (\gamma )$ defined by}
\begin{equation}
\int_{-\infty }^\infty \dots {\rm d}\nu (\gamma )\equiv 
\lim_{R\rightarrow \infty }\frac{1}{2R}\int_{-R }^R\dots {\rm d}\gamma .
\end{equation} 
{\bf Proof.} 
Looking for a function of the form $k(J)=N(J)^2\rho (J)$ we get
\[ \int_0^\infty k(J)\, {\rm d}J \int_{-\infty }^\infty |J,\gamma \rangle \langle J,\gamma |\, {\rm d}\nu (\gamma )
=\sum_{n<\Lambda _m}\left( \frac{1}{\rho _n}\int_0^\infty J^n\rho (J) {\rm d}J \right)|n\rangle \langle n|. \]
In the case $\sigma ''=0$ we  can choose $\rho (J)=-\frac{1}{\alpha }{\rm e}^{\frac{1}{\alpha }J}$ since
the change of variable $J=-\alpha t$ leads to
\[ \frac{1}{\rho _n}\int_0^\infty J^n\rho (J) {\rm d}J = \frac{1}{n!}\int_0^\infty t^n{\rm e}^{-t}dt =1\qquad
{\rm for\ any}\quad n\in \mathbb{N}. \]
The modified Bessel function 
\begin{equation}
K_\nu (z)=\frac{\pi }{2}\frac{I_{-\nu }(z)-I_\nu (z)}{{\rm sin}\, (\nu \pi )}
\qquad {\rm where}\quad 
I_\nu (z)=\sum_{n=0}^\infty \frac{\left(\frac{1}{2}z\right)^{\nu +2n}}
{n!\, \Gamma (\nu +n+1)}.
\end{equation}
satisfies the relation \cite{BP}
\[ \int_0^\infty 2x^{\eta +\xi }K_{\eta -\xi }(2\sqrt{x})\, x^{n-1}dx=
\Gamma (2\eta +n)\, \Gamma (2\xi +n) \]
which for $x=J$, \ $\eta =\frac{1}{2}$, \ $\xi =m-\frac{\alpha }{2}$ becomes
\begin{equation}\label{mellin1}
2\int_0^\infty J^n\, J^{m-\frac{1+\alpha }{2}}\, K_{\frac{\alpha +1}{2}-m}(2\sqrt{J})\, {\rm d}J
=n!\, \Gamma (n+2m-\alpha ).
\end{equation}
The last formula shows that in the case $\sigma ''=-2$ we can choose \cite{BG,AGMKP}
\[ k(J)=\frac{2}{\Gamma (2m-\alpha )}J^{m-\frac{1+\alpha }{2}}\,  K_{\frac{\alpha +1}{2}-m}(2\sqrt{J}).\]
The Bessel function $J_\nu $ satisfies the relation \cite{GR}
\begin{equation}
\int_0^\infty x^\mu J_\nu (ax){\rm d}x=2^\mu a^{-\mu -1}
\frac{\Gamma \left(\frac{1}{2}+\frac{\nu }{2}+\frac{\mu }{2}\right)}
{\Gamma \left(\frac{1}{2}+\frac{\nu }{2}-\frac{\mu }{2}\right)} 
\end{equation}
for $-{\rm Re}\, \nu -1<{\rm Re}\, \mu <\frac{1}{2}$. If we use the substitutions 
$a=2$, $x=\sqrt{J}$, $\nu =1-\alpha -2m$ and $\mu =\alpha +2m+2n$ then we get 
the relation 
\[ \int_0^\infty J^n\, J^{m-\frac{1-\alpha }{2}}\, J_{1-\alpha -2m}(2\sqrt{J})\, {\rm d}J=
\frac{n!}{\Gamma (1-\alpha -2m-n)} \]
which shows that in the case $\sigma ''=2$ we can choose \cite{RR}
\[ \rho (J)=\Gamma (1-\alpha -2m) J^{m-\frac{1-\alpha }{2}}\, J_{1-\alpha -2m}(2\sqrt{J}).\qquad \opensquare \]
Let $m\in \mathcal{M}$ and
\begin{equation}
\tilde \Lambda _m=\left\{ \begin{array}{ll}
\infty & {\rm if} \ \ \sigma(s)\!=\!s\ \ {\rm or}\ \ \sigma (s)\!=\!1\!-\!s^2\\[2mm]
{\rm sup}\, \mathcal{M}-m &  {\rm if} \ \ \sigma(s)\!=\!s^2\!-\!1\ \ {\rm or}\ \ \sigma (s)\!=\!s^2\!+\!1
\end{array} \right.  
\end{equation}
The functions  
\begin{equation}
|\tilde n\rangle =\tilde \phi _{m+n,m}\qquad {\rm with}\quad 0\leq n<\tilde \Lambda _m
\end{equation}
form an orthonormal system, and satisfy the relation
\begin{equation}
(\tilde H_m-\tilde \lambda _m)|\tilde n\rangle =\tilde e_n|\tilde n\rangle 
\end{equation}
with $\tilde e_n$ defined by 
\begin{equation} \fl
\tilde e_n\!=\!\tilde \lambda _{m+n}\!-\!\tilde \lambda _m
\!=\!\left\{ 
\begin{array}{rcl}
\frac{\delta ^2}{\beta _m^2}\frac{n(n+\beta _m)}{(n+\beta _m/2)^2} & {\rm if} & \sigma (s)=s\\[3mm]
\frac{n(n-2\alpha _m)(n-\alpha _m-{\rm i}\delta /2\alpha _m)(n-\alpha _m+{\rm i}\delta /2\alpha _m)} {(n-\alpha _m)^2}
& {\rm if } & \sigma (s)=1-s^2\\[3mm]
-\frac{n(n-2\alpha '_m)(n-\alpha '_m-\delta /2\alpha '_m)(n-\alpha '_m+\delta /2\alpha '_m)} {(n-\alpha '_m)^2} 
 & {\rm if } & \sigma (s)\!=\!s^2\pm 1.
\end{array} \right.
\end{equation}
where $\beta _m=2m+2\beta -1$, \, $\alpha _m=\frac{1+\alpha }{2}-m$ and $\alpha '_m=\frac{1-\alpha }{2}-m$.\\[5mm]
We define the Gazeau-Klauder coherent states
\begin{equation} \label{coherent}
 |J,\gamma \rangle =N(J)^{-1}\sum_{n<\tilde \Lambda _m}\frac{J^{n/2}}{\sqrt{\tilde \rho _n}}
{\rm e}^{-{\rm i}\tilde e_n\gamma }|\tilde n\rangle 
\end{equation}
where $N(J)$ is a normalizing constant and 
\[  
\begin{array}{l} 
\tilde \rho _n=
\frac{\delta ^{2n}}{\beta _m^{2n}} \ \frac{\Gamma ^2(\beta _m/2+1)}{\Gamma (\beta _m+1)} \ 
\frac{n! \Gamma (n+\beta _m+1) }{ \Gamma ^2(n+\beta _m/2+1) }
\end{array}
\] 
in the case  $\sigma (s)=s$,
\[ \fl 
\begin{array}{l} 
\tilde \rho _n=
\frac{\Gamma ^2(1-\alpha _m) }
{\Gamma (1-2\alpha _m)\, \Gamma (1-\alpha _m-{\rm i}\delta /2\alpha _m)\, \Gamma (1-\alpha _m+{\rm i}\delta /2\alpha _m) }\, 
 \frac{n!\, \Gamma (n-2\alpha _m+1)\, \Gamma (n-\alpha _m-{\rm i}\delta /2\alpha _m+1)\, 
\Gamma (n-\alpha _m+{\rm i}\delta /2\alpha _m+1) } { \Gamma ^2(n-\alpha _m+1) } 
\end{array} \]
in the case $\sigma (s)=1-s^2$, and 
\[  \fl 
\begin{array}{l} 
\tilde \rho _n\!=\!
\frac{(-1)^n\, \Gamma ^2(1-\alpha '_m) }
{\Gamma (1-2\alpha '_m)\, \Gamma (1-\alpha '_m-\delta /2\alpha '_m)\, \Gamma (1-\alpha '_m+\delta /2\alpha '_m) }\, 
 \frac{n!\, \Gamma (n-2\alpha '_m+1)\, \Gamma (n-\alpha '_m-\delta /2\alpha '_m+1)\, 
\Gamma (n-\alpha '_m+\delta /2\alpha '_m+1) } { \Gamma ^2(n-\alpha '_m+1) } 
\end{array}
\]
in the cases $\sigma (s)=s^2\pm 1$. One can remark that 
\[ N(J)^2=_2\!\!F_1\left(1+\frac{\beta _m}{2},1+\frac{\beta _m}{2};1+\beta _m; \frac{\beta _m^2}{\delta ^2}J\right) \]
in the case  $\sigma (s)=s$, 
\[ N(J)^2=_2\!\!F_3\left(1\!-\!\alpha _m,1\!-\!\alpha _m;1\!-\!2\alpha _m,1\!-\!\alpha _m\!-\! \frac{\delta {\rm i} }{2\alpha _m}, 
1\!-\!\alpha _m\!+\!\frac{\delta {\rm i} }{2\alpha _m};J\right) \]
in the case  $\sigma (s)=1-s^2$, and $N(J)^2$ is a finite sum in the cases $\sigma (s)=s^2\pm 1$.
If $\sigma (s)=s$ then the coherent states (\ref{coherent}) can be defined only for 
$|J|<\frac{\delta ^2}{\beta _m^2}$.\\[5mm]
{\bf Theorem 13.}
{\it If $\sigma (s)\!=\!1\!-\!s^2$ or $\sigma (s)\!=\!s^2\pm 1$ then there exists a real function $k(J)$ such that
\begin{equation}
\int_0^\infty k(J)\, {\rm d}J \int_{-\infty }^\infty |J,\gamma \rangle \langle J,\gamma |\, {\rm d}\nu (\gamma )
=\sum_{n<\tilde \Lambda _m}|\tilde n\rangle \langle \tilde n|=\mathbb{I} 
\end{equation}
with ${\rm d}\nu (\gamma )$ defined by}
\begin{equation}
\int_{-\infty }^\infty \dots {\rm d}\nu (\gamma )\equiv 
\lim_{R\rightarrow \infty }\frac{1}{2R}\int_{-R }^R\dots {\rm d}\gamma .
\end{equation}
{\bf Proof.}
Looking for a function of the form $k(J)=N(J)^2\rho (J)$ we get
\[ \int_0^\infty k(J)\, {\rm d}J \int_{-\infty }^\infty |J,\gamma \rangle \langle J,\gamma |\, {\rm d}\nu (\gamma )
=\sum_{n<\tilde \Lambda _m}\left( \frac{1}{\tilde \rho _n} \int_0^\infty J^n\rho (J) {\rm d}J \right)|\tilde n\rangle \langle \tilde n|. \]
In the case $\eta =1$, $m=q$, $n=0$, the relation involving the Meijer's $G$-function \cite{L}
\begin{equation} \fl 
\int_0^\infty z^{s-1}\ G_{p,q}^{m,n}\!\left. \left(\begin{array}{l}
a_1,...,a_p\\
b_1,...,b_q
\end{array} \right| \eta z\right) {\rm d}z=\frac{\eta ^{-s}\prod_{j=1}^m\Gamma (b_j+s)\prod_{j=1}^n\Gamma (1-a_j-s)}
{\prod_{j=m+1}^q\Gamma (1-b_j-s)\prod_{j=n+1}^p\Gamma (a_j+s)}
\end{equation}
becomes 
\begin{equation}
\int_0^\infty z^{s-1}\ G_{p,q}^{q,0} \left. \left(\begin{array}{l}
a_1,...,a_p\\
b_1,...,b_q
\end{array} \right| z\right) {\rm d}z=\frac{\Gamma (b_1+s)...\Gamma (b_q+s)}{\Gamma (a_1+s)...\Gamma(a_p+s)}.
\end{equation}
In order to get $\int_0^\infty J^n\rho (J) {\rm d}J=\tilde \rho _n$ we choose \cite{CF}
\[  \fl \begin{array}{l}
\rho(J)\!= \!\frac{\Gamma ^2(1-\alpha _m) }
{\Gamma (1-2\alpha _m)\, \Gamma (1-\alpha _m-{\rm i}\delta /2\alpha _m)\, \Gamma (1-\alpha _m+{\rm i}\delta /2\alpha _m) }\, 
G_{2,4}^{4,0}\left. \!\!\left( \!\!\begin{array}{l}
-\alpha _m,-\alpha _m\\
0, -2\alpha _m, -\alpha _m\!-\!{\rm i}\delta /2\alpha _m, -\alpha _m\!+\!{\rm i}\delta /2\alpha _m 
\end{array} \right| J\right)
\end{array} 
\]
in the case $\sigma (s)=1-s^2$, and
\[ \fl  \begin{array}{l}
\rho(J)\!= \!\frac{\Gamma ^2(1-\alpha '_m) }
{\Gamma (1-2\alpha '_m)\, \Gamma (1-\alpha '_m-\delta /2\alpha '_m)\, \Gamma (1-\alpha '_m+\delta /2\alpha '_m) }\, 
G_{2,4}^{4,0}\left. \!\!\left( \!\!\begin{array}{l}
-\alpha '_m,-\alpha '_m\\
0, -2\alpha '_m, -\alpha '_m\!-\!\delta /2\alpha '_m, -\alpha '_m\!+\!\delta /2\alpha '_m 
\end{array} \right| J\right)
\end{array} 
\] 
in the cases $\sigma (s)=s^2\pm 1$.$\qquad \opensquare $
\section{Concluding remarks}

If we use in equation ${H}_m \Phi _{l,m}=\lambda _l\Phi _{l,m}$ a change of variable 
$(a',b')\longrightarrow (a,b):x\mapsto s(x)$ 
such that $ds/dx=\kappa (s(x))$ or $ds/dx=-\kappa (s(x))$ and
define the new functions 
\begin{equation}
\Psi _{l,m}(x)=\sqrt{\kappa (s(x))\, \varrho (s(x))}\, \Phi _{l,m}(s(x))
\end{equation}
then we get an equation of Schr\" odinger type 
\begin{equation}
-\frac{d^2}{dx^2}\Psi _{l,m}(x)+V_m(x)\Psi _{l,m}(x)
=\lambda _l\Psi _{l,m}(x) .\label{Schrod}
\end{equation}
Since
\[  \fl   \int_{a'}^{b'} \Psi _{l,m}(x)\Psi _{k,m}(x) dx=
\int_{a'}^{b'}\Phi _{l,m}(s(x))\Phi _{k,m}(s(x))\varrho (s(x))\frac{d}{dx}s(x)dx
=\int_a^b\Phi _{l,m}(s)\Phi _{k,m}(s)\varrho (s)ds \]
the functions $\Psi _{l,m}(x)$ are square integrable (resp. orthogonal) if and only if the 
corresponding functions $\Phi _{l,m}(s)$ are square integrable (resp. orthogonal).

If $ds/dx=\kappa (s(x))$ then the operators corresponding to $A_m$ and $A_m^+ $ are
\begin{equation}\label{tildeA+}
 \begin{array}{l}
{\mathcal A}_m=[\kappa (s)\varrho (s)]^{1/2}A_m
[\kappa (s)\varrho (s)]^{-1/2}|_{s=s(x)}
=\frac{d}{dx}+W_m(x)\\[2mm]
{\mathcal A}_m^+ =[\kappa (s)\varrho (s)]^{1/2}A_m^+ 
[\kappa (s)\varrho (s)]^{-1/2}|_{s=s(x)}
=-\frac{d}{dx}+W_m(x)
\end{array}
\end{equation}
where the {\em superpotential} $W_m(x)$ is given by the formula 
\begin{equation}\label{Wm}
W_m(x)=-\frac{\tau (s(x))}{2\kappa (s(x))}
-\frac{2m-1}{2\kappa (s(x))}\frac{d}{dx}\kappa (s(x))\, .
\end{equation}

We have 
\begin{equation}
  {\mathcal A}_m\Psi _{l,m}(x)=\Psi _{l,m+1}(x) \qquad 
{\mathcal A}_m^+ \Psi _{l,m+1}(x)=(\lambda _l-\lambda _m)\Psi _{l,m}(x)
\end{equation}
\begin{equation}\label{factor} \fl 
   -\frac{d^2}{dx^2}+V_m(x)-\lambda _m={\mathcal A}_m^+ {\mathcal A}_m\qquad
-\frac{d^2}{dx^2}+V_{m+1}(x)-\lambda _m={\mathcal A}_m{\mathcal A}_m^+ 
\end{equation}
and
\begin{equation}  
\Psi _{l,m}(x)=
\frac{{\mathcal A}_m^+ }{\lambda _l-\lambda _m}
\frac{{\mathcal A}_{m+1}^+ }{\lambda _l-\lambda _{m+1}}...
\frac{{\mathcal A}_{l-2}^+ }{\lambda _l-\lambda _{l-2}}
\frac{{\mathcal A}_{l-1}^+ }{\lambda _l-\lambda _{l-1}}
\Psi _{l,l}(x)
\end{equation} 
for each $l\in \{ 0,1,...,\nu \}$ and each $m\in \{ 0,1,...,l-1\}$.

If we choose the change of variable $s=s(x)$ such that 
$ds/dx=-\kappa (s(x))$, then 
\begin{equation}
{\cal A}_m=-\frac{d}{dx}+W_m(x)\qquad 
{\cal A}_m^+ =\frac{d}{dx}+W_m(x)
\end{equation}
and
\begin{equation}
W_m(x)=-\frac{\tau (s(x))}{2\kappa (s(x))}
+\frac{2m-1}{2\kappa (s(x))}\frac{d}{dx}\kappa (s(x)).
\end{equation}
Some very similar results can be obtained in the case of operators $\tilde H_m$.
The Gazeau-Klauder systems of coherent states defined in the previous section correspond through
the considered change of variables to some systems of coherent states useful in quantum mechanics.
\section*{Acknowledgment}
The author gratefully acknowledges the support provided by CNCSIS under the grant IDEI 992 - no. 31/2007.

\end{document}